\documentclass[sigplan,nonacm]{acmart}
\AtBeginDocument{%
  }

\usepackage[dvipsnames]{xcolor}
\usepackage{cathywumath}
\usepackage{hyperref}
\usepackage[short,12hr]{datetime}
\usepackage{pgfplots}
\usepackage{wrapfig}
\usepackage{hyphenat}
\usepackage{cuted}
\usepackage{fancyvrb}
\usepackage{cancel}
\usepackage{mdframed}
\newif\ifdraft
\drafttrue

\newcommand{\eat}[1]{}

\makeatletter
\renewcommand{\listoftodos}[1][\@todonotes@todolistname]{\@starttoc{tdo}{#1}}
\makeatother

\setcopyright{none}
\acmConference{}{}{}{}
\acmBooktitle{}
\begin{document}

\title{Proof Interfaces for Exploratory Mathematics}

\author{Nishant Kheterpal}
\email{nskh@umich.edu}
\author{Matthew Keenan}
\email{mckeenan@umich.edu}
\author{Cyrus Omar}
\email{comar@umich.edu}
\author{Jean-Baptiste Jeannin}
\email{jeannin@umich.edu}
\affiliation{%
  \institution{University of Michigan}
  \city{Ann Arbor, MI}
  \country{USA}
}

\renewcommand{\shortauthors}{Kheterpal et al.}

\begin{abstract}
  This paper describes a mathematics interface intended for both educational applications and exploratory mathematics.
  Both learners and mathematics practitioners often use pen and paper or a whiteboard to perform equational reasoning, manipulate expressions, or construct proofs. These workflows lead to a variety of problems: transcription errors, tedious writing and notation, and/or unclear standards for proof justification.
  We extend the \textbf{Hazel Prover}, an equational-reasoning interface in the Hazel live programming environment, to add capabilities for exploratory mathematics, with varying levels of verbosity and automation aimed at both students and expert users.
  Students require more deliberate practice when learning mathematical concepts and correspondingly more verbose justifications, while experts may benefit from significant mathematical automation.
  With this in mind, our interface supports multiple levels of mathematical automation and simplification, grounded in a rewrite search architecture.
  For expert users, motivated by a gap in the accessibility of formal methods, we support proof export to the Rocq theorem prover, extending this rewrite search to proof tactics.
  This paper closes with several case studies covering elementary- to college-level mathematics.
\end{abstract}
\maketitle

\section{Introduction}
\label{sec:introduction}

Both learners and mathematics practitioners conduct ``exploratory math'': the kinds of open-ended rewriting of terms and bounding of quantities that are often done on a whiteboard.
This whiteboard work, while frequently useful for creative purposes, suffers from significant friction: rewriting terms is tedious, and small errors (flipping $+$ to $-$, for example) are easily made and hard to notice.
Results need to be preserved or shared, by typesetting the derivation in \LaTeX{}, copying it onto paper for submission, or photographing the whiteboard.

Expert users may conduct math directly in \LaTeX{}, where copying lines is trivial but the syntax may become tedious, especially for constructs like matrices.
Meanwhile, computer algebra systems such as Mathematica~\cite{Mathematica}, SymPy~\cite{SymPyPaper}, Maple~\cite{maple_2026}, or MATLAB~\cite{matlab_symbolic_math_toolbox_2025} work well to solve expressions or perform fixed, known transformations on expressions or equations, but do not permit the open-ended exploration needed to bound terms or prove theorems.

Students, on the other hand, require more deliberate practice when learning mathematical concepts and correspondingly more verbose justifications and less automation than experts.
These students may still struggle, in particular when learning formal proof, to break down an argument into sufficiently small steps, or to apply new rules correctly \cite{winer2022proof}.
An instructor might structure practice by limiting the available rules or controlling step granularity.

Nonetheless, both groups of users could benefit from a ``digital whiteboard'' that adds structure to their work.
For experts, this interface must permit open-ended exploration while ensuring incremental correctness.
Students, on the other hand, require less automation but a different sort of structure: guidance and scaffolding to help practice mathematical concepts while abstracting away others.
These needs motivate varying the mathematical capabilities, automation, and simplification the interface provides.

Full-scale proof assistants offer industry-standard verification capabilities but can be difficult to use for non-experts. Rocq \cite{rocq_prover_2025}, Lean \cite{leanprover}, and Isabelle \cite{nipkow_isabellehol_2002} have been used in educational settings and to formalize mathematics. However, the learning curve and syntax can challenge even experts. Educational proof assistants with simplified interfaces and abstractions have been built, such as Waterproof, based on Rocq, which has been used to teach real analysis in the past \cite{wemmenhove2022waterproof,wemmenhove2025waterproof}.
Similar systems have been built based on Lean \cite{massot2024teaching,minh2025lean}. Graphical tools such as Edukera teach proofs visually \cite{rognier_presentation_2016}. Tools aimed beyond proofs such as GeoGebra \cite{hohenwarter_dynamic_2007}, Graspable Math \cite{ottmar_graspable_2015}, and MathXPert \cite{beeson_design_1998} inspired parts of our design.

In this work, we present an extension of the \textit{Hazel Prover}, an existing equational reasoning interface, to support exploratory mathematics.
Past work using the Hazel Prover focused on the development and deployment of the prover in classrooms for teaching structural induction proofs \cite{keenan2026prover}.
The Hazel Prover is built in Hazel, a live programming environment that supports direct manipulation of program structure \cite{omar_live_2019}.
The existing educational uses of the Hazel Prover led to a natural second question: can a powerful exploratory interface also be used to teach more basic mathematics?
Similar levels of automation in the broader Hazel Prover were found to be valuable for teaching in deployments in college-level programming languages courses \cite{keenan2026prover}.

For practitioners in applied fields such as aerospace engineering and physics, this work also connects to the accessibility of formal methods.
These practitioners regularly engage in rigorous mathematical reasoning --- deriving equations of motion, linearizing dynamics, and discretizing continuous systems --- yet formal methods training is not part of their background.
We aim to lower the barrier to using formal methods for symbolic reasoning by developing a platform in which users can manipulate equations or expressions and emit a formal proof of their steps at the end.

Our contributions are the design and implementation of an interface for exploratory mathematics; a customizable rewrite search architecture (Section~\ref{sec:architecture}) supporting mathematical capabilities organized by educational level and automation ranging from single steps to simplification; formal proof export from step-by-step derivations; and case studies investigating suitability for tasks from order of operations through algebra and calculus to exploratory numerical methods (Section~\ref{sec:case-studies}).
Section~\ref{sec:design-criteria} describes the interface's design goals.

\vspace{-0.5\baselineskip}
\section{Design Goals}
\label{sec:design-criteria}

In order to ensure our interface is useful for both students and experts, it should fit existing workflows (either for learning exercises or exploration) and be straightforward to use.
To guide our design, we established a set of design criteria, informed by our past work on the Hazel Prover \cite[Section 2]{keenan2026prover}.
Our criteria for usability and workflow fit are similar to design goals in that work, but our goals around coverage and proof export are targeted at the needs of numerical methods practitioners and other experts, rather than students.

Our design goals are:
First, \emph{low setup cost}: a new user should be able to begin solving representative tasks with little configuration or theorem-prover-specific training.
Second, \emph{low interaction overhead}: routine manipulations such as rewrites, differentiation, and substitution should not require too many actions.
Third, \emph{step granularity}: users should be able to take steps that are neither too large (simplifying a full expression) nor too small (only commuting terms).
Fourth, \emph{mathematical coverage}: the interface should support the symbolic tasks in our case studies (Section~\ref{sec:case-studies}).

\eat{
  These criteria motivate the following research questions:
  \begin{quote}
    {\bf RQ1 (Usability):} Can engineering students and practitioners learn to use the interface quickly enough to complete representative exploratory-mathematics tasks with minimal instruction?
  \end{quote}
  \begin{quote}
    {\bf RQ2 (Workflow fit):} Does the interface reduce the bookkeeping burden of multi-step symbolic derivations while preserving visibility into intermediate reasoning steps?
  \end{quote}
  \begin{quote}
    {\bf RQ3 (Coverage):} Which classes of engineering mathematics tasks are well supported by the interface, and where do users encounter missing operations or breakdowns in expressiveness?
  \end{quote}
  \begin{quote}
    {\bf RQ4 (Trust and verification):} Do users understand and value the interface's verification and proof-export mechanisms, and do these mechanisms help them detect or avoid derivation errors?
  \end{quote}
  \todo{Cyrus: These research questions promise user evidence not yet reported here; either soften them to case-study questions or add an expert walkthrough/user study.}
  \todo{Tanner: Revisit these research questions in light of the participant-experiment scope change. State clearly which questions the case-study evaluation answers, and revise, move, or remove questions that are no longer answered.}
}

\vspace{-0.5\baselineskip}
\section{Architecture}
\label{sec:architecture}

Once an expression has been entered, users begin each step by selecting an expression or subexpression and opening a dropdown menu.
In this dropdown, experts might seek a useful rewrite or simplification, or students might be asked to apply specific rules, such as distribution or trigonometric identities.
The interface supports applying or proposing transformations, checking their permissibility, and recording accepted derivations for proof export. Upon applying or proposing a valid rewrite, the user may take another step.

For example, consider the expression $(2x-3)(x+4)$.
In a student-facing
exercise, Hazel might present distribution as the next available rule.
An instructor might, at first, permit students learning binomial expansion to apply only one simple distribution, producing $(2x-3)\cdot x + (2x-3)\cdot 4$.
But a later lesson might have students do more: $2x^2 - 3x + 8x - 12$, or perhaps a version with like terms collected: $2x^2 + 5x - 12$.
An expert user would find this tedious and might prefer simply to click a button that says ``Simplify''.
These uses differ in how much work Hazel performs, but both require the same basic guarantee: the accepted result must be correct, and an accepted derivation may be exported to Rocq.
In this section, we detail an architecture that supports both student and expert users, allows for a range of automation, is customizable and extensible, and permits export to a theorem prover.

\subsection{Rewrite Search}
\label{sec:rewrite-search}

\begin{figure}
       \centering
       \includegraphics[width=0.7\linewidth]{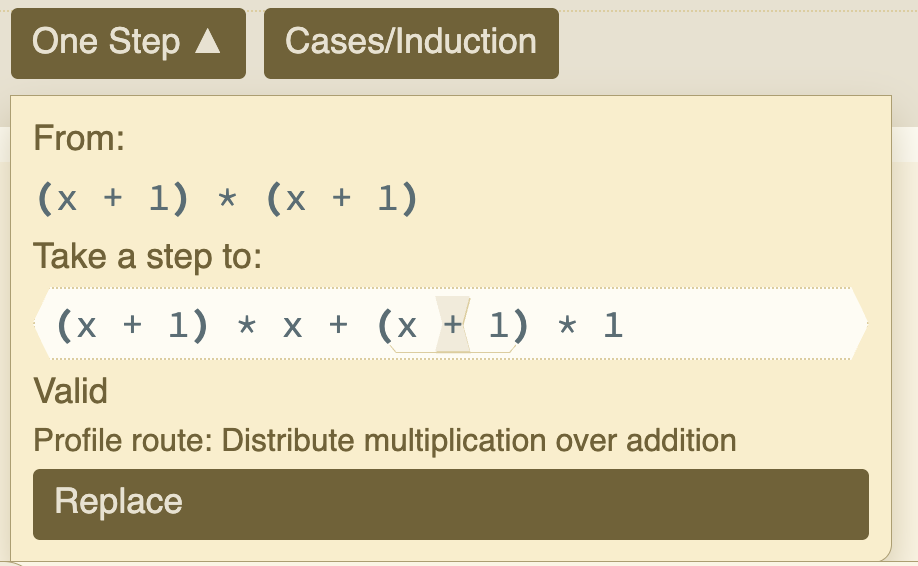}
       \caption{One Step permits only individual rewrites.}
       \label{fig:one-step-mode}
       \Description{A Hazel Prover panel in One Step mode
              showing distribution of a binomial.}
\end{figure}

Our interface has three automation modes: One Step, Check Result, and Suggest Result, which we detail below.
One Step (Figure~\ref{fig:one-step-mode}) is the most restricted mode, in which users must take a single step at a time: for example,
distributing $(x+1)(x+1)$ to $x^2 + x + x + 1$.
The nature of what constitutes
a ``step'' is defined by the current math profile (see
Section~\ref{sec:math-profiles}).
For example, this distribution could
be more verbose: $(x+1)\cdot x + (x+1)\cdot 1$, or less verbose: $x^2 + 2x + 1$.
One Step does not merely require that the proposed result be mathematically
correct: it must be reachable by one enabled rewrite.
Thus, evaluating
$3+4\cdot4$ directly to $19$ is rejected because it combines multiplication
and addition, whereas evaluating the selected subexpression $4\cdot4$ to $16$
is accepted.

\begin{figure}
       \centering
       \includegraphics[width=0.7\linewidth]{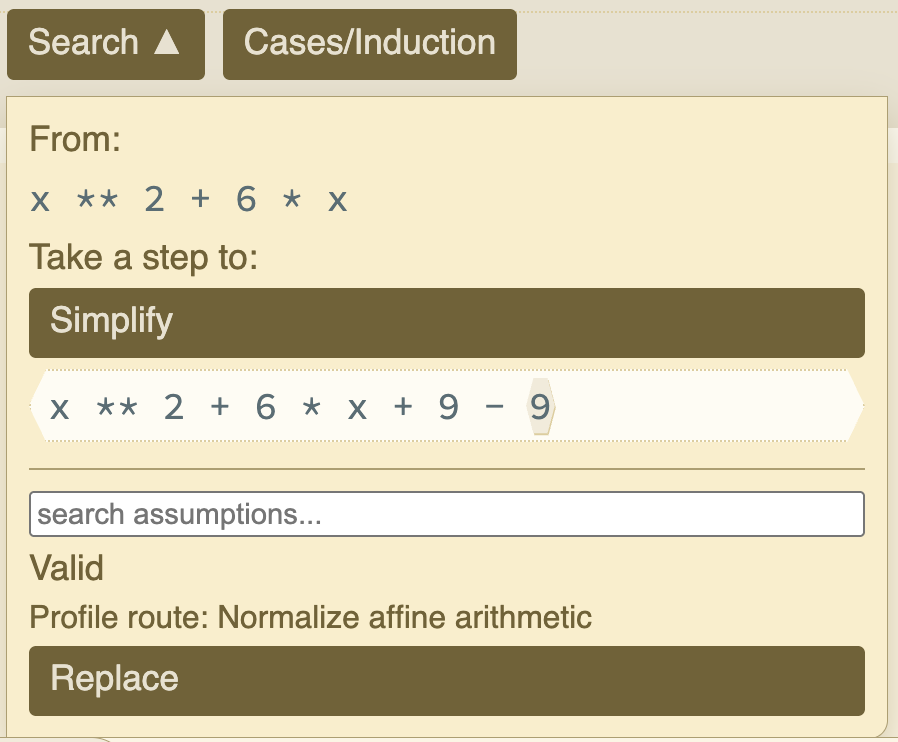}
       \caption{In Check Result and Suggest Result modes, the interface checks the validity of a suggestion produced by a validated sequence of rewrites. \vspace{-1em}}
       \label{fig:check-result-mode}
       \Description{A Hazel Prover panel in Check Result mode
              showing simplification of a quadratic expression, with the profile route
              ``Normalize affine arithmetic.''}
\end{figure}

The Check Result and Suggest Result modes (Figure~\ref{fig:check-result-mode}) are aimed more at expert users,
with Suggest Result extending Check Result to automatically simplify the
selected subexpression.
In Check Result mode, the user proposes an equivalent result that may require multiple rewrites, and the interface checks its validity.
Validity depends on the current math profile.
For example, a user may
select $(2x-3)(x+4)$ and enter $2x^2+5x-12$.
Check Result may accept this as one
interaction, but only if the active profile permits the distribution and cleanup
needed to reach that result.
If collecting like terms is disabled, mathematical equivalence alone does not make the result acceptable.

\eat{
       Check Result takes the selected subexpression, proposed result, and active profile, and searches the graph of enabled rewrites to determine whether the
       result is reachable.
       The approach uses depth-limited search over the graph
       induced by the rewrites in the profile.
       Certain specialized checks are performed
       for common operations such as arithmetic evaluation and expanding polynomials,
       and expressions are normalized to prevent the search from considering arbitrary
       associations and commutations of sums and products.
       Thus, the search is not a
       general equivalence oracle: it returns either a permitted rewrite sequence or
       failure, and is deliberately restricted to the work that the active profile
       permits.}

\subsection{Math Profiles and Automation Policy}
\label{sec:math-profiles}

Exercises are associated with a \emph{math profile}, which defines the enabled
rewrites and automation.
Those rewrites are suggested to
the user contextually: after a user selects a subexpression, Hazel offers
identities and rules whose patterns match that selection.
For example, a rule
for $\sin^2(\cdot)$ is offered only when the selected term has that form.
Users choose among applicable suggestions.
The same profile bounds Check Result search: Hazel accepts a result only if a sequence of permitted rewrites reaches it.

Our design is oriented around the principle that a student who has moved on to a
later mathematical level has mastered and can abstract away more primitive
skills.
For example, a student learning trigonometry need not verbosely compute
arithmetic according to the order of operations or manually distribute
binomials.
As such, we envision progressing through math levels iteratively: first increasing automation (algebraic simplification, factoring, etc.) as skills are learned, and then ``unlocking'' new mathematical capabilities when introducing a new subject such as trigonometry.

\subsection{Proof Export}

A completed sequence of rewrite steps in the Hazel interface constitutes a
structured but informal proof.
We export these step sequences to the Rocq
theorem prover for machine-checked verification \cite{rocq_prover_2025}.
An
exported proof uses the rewrite rules from each step, either taken directly in
One Step mode or found by search in more automated modes.
The Rocq proof
assistant is integrated in two ways: first, in the browser, where JSCoq (a web interface for the Rocq theorem prover) is used
to check the validity of each step, and second, through an export feature that generates a
complete proof script corresponding to all steps taken
\cite{gallego_arias_jscoq_2017_architecture}. In Check Result and Suggest Result
modes, after a rewrite sequence has been found using search, the interface
generates a Rocq theorem and sends it to a background JSCoq worker to check.

\eat{
       More powerful Rocq tactics than simple rewrites can be used to check step correctness because, once JSCoq is invoked, Hazel has already found a sequence that respects the math profile.
       Correspondingly, automated Rocq tactics such as \texttt{lra}, \texttt{ring}, and \texttt{field}
       can discharge the proof obligations of each step though they may include automations beyond the scope of an active math profile.
       In our experience, attempting to perform search using the current math profile's enabled rewrites, JSCoq, and the Coq tactic language Ltac~\cite{delahaye_ltac_2000} was too slow to be practical.
       Hazel therefore performs the profile-bounded search,
       while JSCoq and Rocq check the resulting obligations and support complete proof
       export.
       If Hazel cannot find a rewrite sequence, an equivalence check may still
       be performed within Rocq, but is reported as ``Equivalent, outside profile'':
       Rocq can establish equality, but Hazel does not treat it as an allowed step.
}

\section{Extended Case Studies}
\label{sec:case-studies}

This section presents several case studies of increasing complexity and evaluates our interface against them.
The case studies are inspired by courses spanning algebra education through calculus, and are ordered so that each builds on capabilities established by the prior one.
The simplest case studies are intended to be valuable teaching tools where proof export is trivial using automated tactics; our later case studies aim to simplify formal proof for numerical methods practitioners. Due to space constraints, only select case studies are described below. In each case study, we briefly comment on the experience of using the interface and the extent to which it meets our design criteria (Section~\ref{sec:design-criteria}). Our design criteria about \textit{low setup cost} are consistently met across case studies: the exercises are browser-based and interface elements to take rewrites are presented upon selecting terms.

\begin{figure}
  \centering
  \Description{A completed Hazel Prover proof of $x^2+6x+5=(x+3)^2-4$, with steps labeled affine normalization, polynomial expansion, profile cleanup, and reflexivity.}
  \includegraphics[width=\columnwidth]{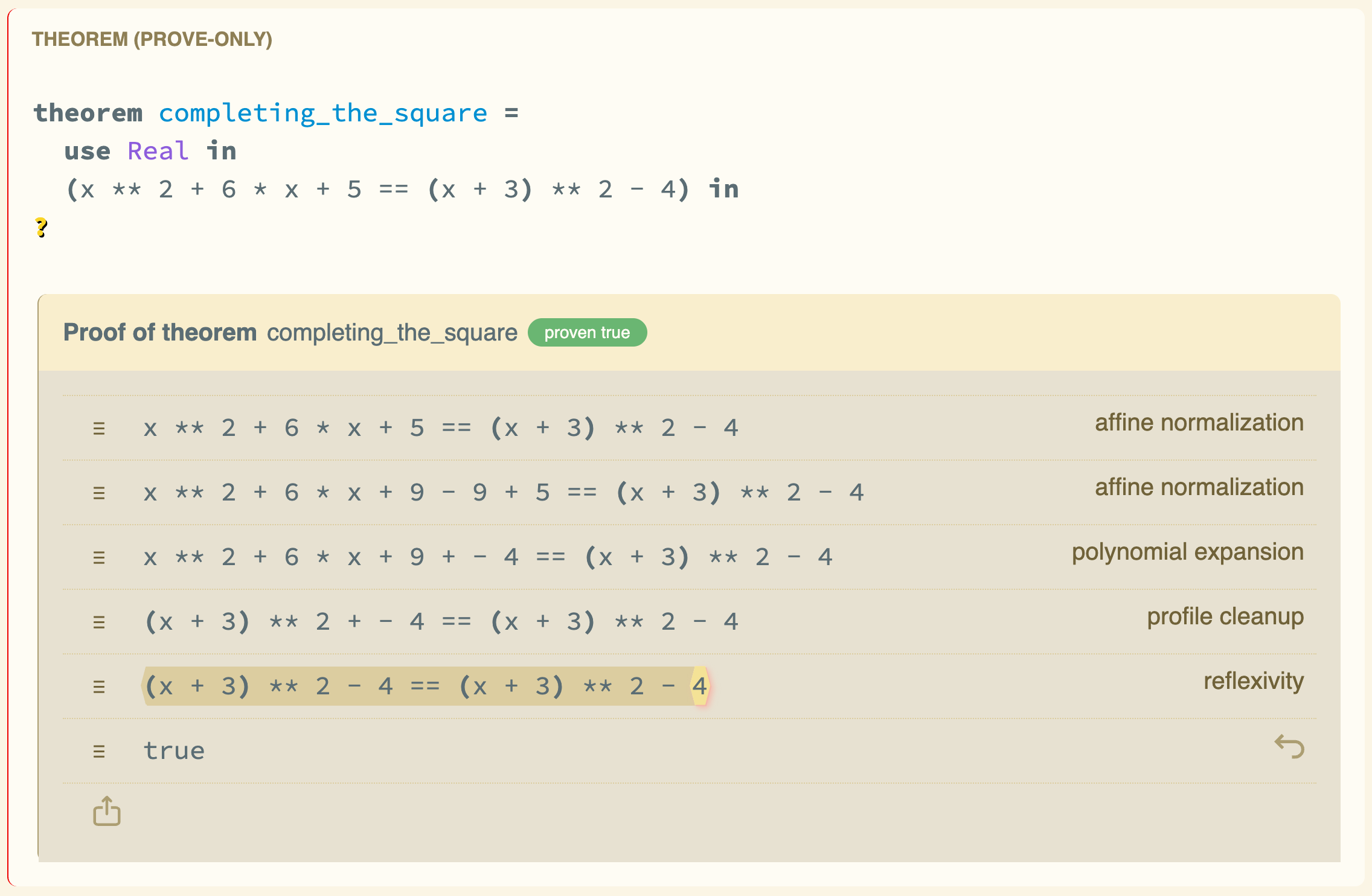}
  \caption{Completing the square in Hazel Prover. \vspace{-1em}}
  \label{fig:completing-the-square}
\end{figure}

\textbf{Order of Operations.}
The first case study asks students to simplify the arithmetic expression $3+4\cdot2^2-\frac{6}{3}$ step by step to obtain $17$, while obeying the order of operations.
It uses One Step, the most verbose mode.

The verbose steps in this case study are by design, given the target student audience. In this way, our goals around \textit{low interaction overhead} and \textit{step granularity} are met: users must take small steps, each corresponding to one arithmetic operation, in a sequence that obeys the order of operations. The \textit{mathematical coverage} goal is trivially met by built-in arithmetic operations.

\eat{
  \paragraph{FOIL, Written Out.}
  The first binomial-expansion case study asks students to expand $(x+1)(x+1)$ into the four explicit products $x\cdot x+x\cdot1+1\cdot x+1\cdot1$.
  It emphasizes repeated distribution rather than collection of like terms.
  We target this case study at algebra education at the middle or high school level.

  \paragraph{FOIL with Algebraic Cleanup.}
  The second binomial-expansion case study asks students to expand and simplify $(2x-3)(x+4)$ to $2x^2+5x-12$.
  It permits the algebraic cleanup needed to collect the result.
  Like the previous case study, it is targeted at middle or high school algebra.
  Figure~\ref{fig:foil-with-algebraic-cleanup} shows the resulting proof trace,
  including the polynomial-expansion and collection steps recorded by the interface.
}

\eat{
  \begin{figure*}[htbp]
    \centering
    \Description{A Hazel Prover proof trace for $(2x-3)(x+4)=2x^2+5x-12$.
      The displayed steps are polynomial expansion, collection of like terms, and reflexivity.}
    \includegraphics[width=\textwidth]{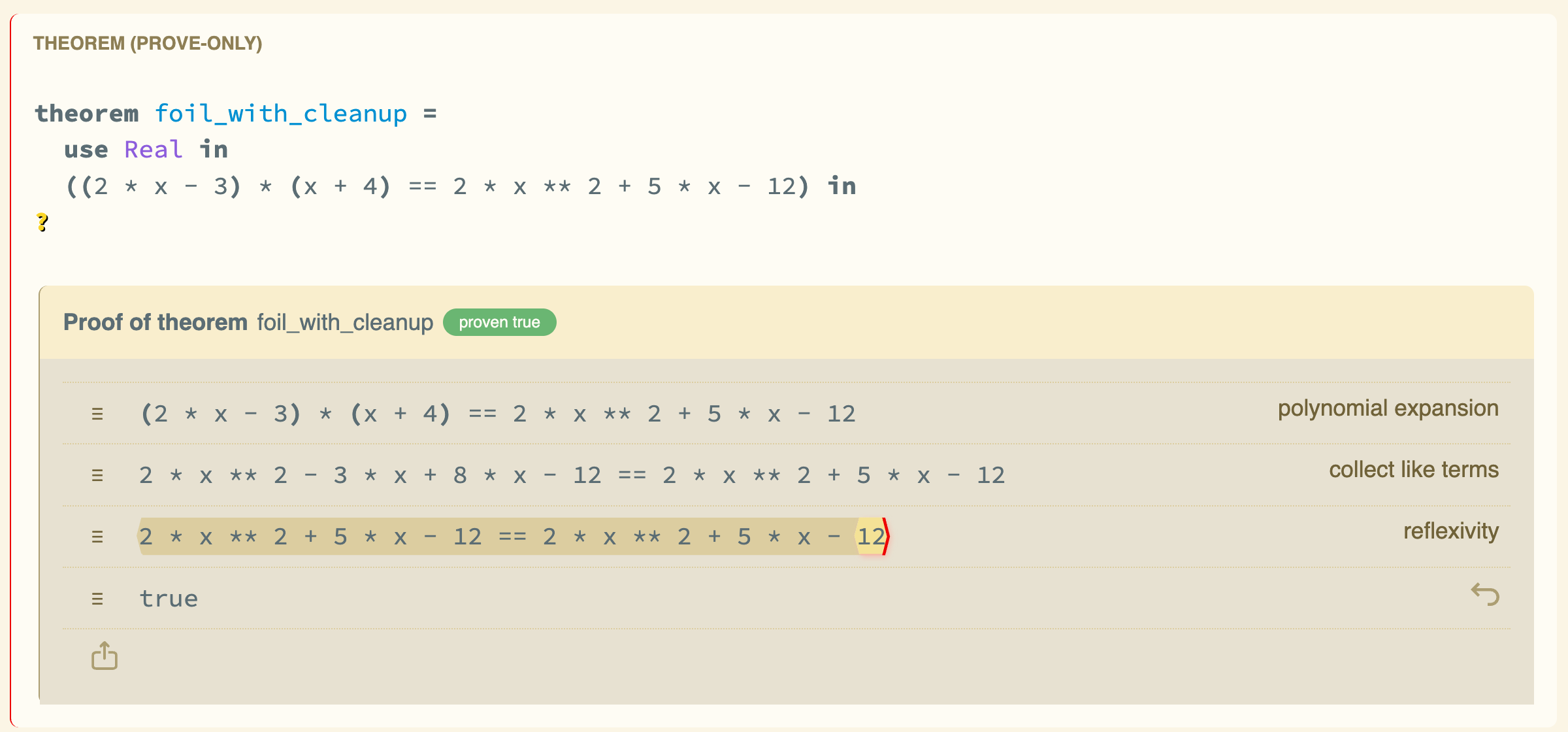}
    \caption{Hazel Prover proof trace for the FOIL-with-algebraic-cleanup case study.
      The proof expands $(2x-3)(x+4)$, collects like terms, and closes by reflexivity.}
    \label{fig:foil-with-algebraic-cleanup}
  \end{figure*}
}

\textbf{Completing the Square.}
Targeted at algebra courses (Figure~\ref{fig:completing-the-square}), users rewrite $x^2+6x+5$ as $(x+3)^2-4$ by completing the square.
Users select $x^2+6x$ and rewrite it as $x^2 + 6x + 9 - 9$, simplify $-9+5$ to $-4$, and factor $x^2 + 6x + 9$ as $(x+3)^2$.
Cleanup and reflexivity close the proof.

This case study is an example of where an expression may need to become more complex (adding $+9 - 9$, here) in order to complete the square. In Check Result mode, this case study permits that exact operation to be taken, and a second step to factor the square $x^2 + 6x + 9$ as $(x+3)^2$. The steps match work on paper, meeting our goals of \textit{interaction overhead}, \textit{step granularity}, and \textit{mathematical coverage}.

\textbf{Trigonometric Identities.}
This case study targets rewriting trigonometric expressions, specifically showing
\[
  1 + 2\sin^4(x) = \frac{7}{4} - \cos(2x) + \frac{1}{4}\cos(4x).
\]

Trigonometric identities introduce a challenge absent from algebraic manipulation: at many intermediate steps, multiple valid rewrites apply to the same subexpression.
For instance, $\sin^2(x)$ can be rewritten using the Pythagorean identity or a double-angle formula; our interface presents both.
This case study targets both precalculus education and numerical methods practitioners who routinely manipulate trigonometric expressions in derivations but lack tooling that records and verifies each step.

Automatic suggestion and application of applicable identities from our interface, combined with simplification of the resulting double-angle expression, reduces the tedium of finding and simplifying trigonometric identities. In this case, our interface reduces \textit{step granularity} in a desirable way compared to pen-and-paper work, with automatic suggestions keeping \textit{interaction overhead} low.

\eat{\paragraph{Symbolic differentiation.}
  This case study moves from equational rewriting to operations on expressions: computing $\frac{d}{dx}[x^3+2x]=3x^2+2$ by applying differentiation rules.
  Derivative rules like linearity, the power rule, and product rule are presented similarly to the trigonometric identities in the previous case study, and the user can select which rule to apply to a selected subexpression.
  This case study builds toward the matrix-valued differentiation required for Jacobian computation and targets introductory calculus and beyond.
}

\textbf{Quadratic Taylor Approximation.}
This case study constructs the (exact) second-order Taylor polynomial of $f(t)=t^2+3t+2$ about $t=0$.
A series of dependent \texttt{let} bindings defines the function and its first two derivatives: \texttt{f1} differentiates \texttt{f}, and \texttt{f2} in turn differentiates \texttt{f1}.
Users first differentiate $f$ and bind the result $f_1(t)=2t+3$.
The interface then substitutes this computed function for $f_1$ in the next derivative expression and differentiates it again, producing $f_2(t)=2$.
Users reuse these results, evaluate the derivatives at a given point, and assemble the finite Taylor polynomial.
This case study targets college-level Calculus II.

In this case study and the next, we encounter limitations of the interface. At the moment, only whole ``programs'' can be stepped, not individual subexpressions. Therefore, steps such as computing the first derivative \texttt{f1} require stepping a larger program that computes \texttt{f2} and the Taylor approximation. In this specific polynomial example, the step burden remains small, as further derivatives reduce in degree towards constants. However, for more complex functions, the step burden can be significant, like in the following case study. As such, \textit{interaction overhead} becomes a concern, though \textit{step granularity} still permits computing derivatives of interest at each step. In this specific case, \textit{mathematical coverage} is as desired, but the following case study lays out shortcomings.

\eat{
  \begin{figure*}[htbp]
    \centering
    \includegraphics[width=0.65\textwidth]{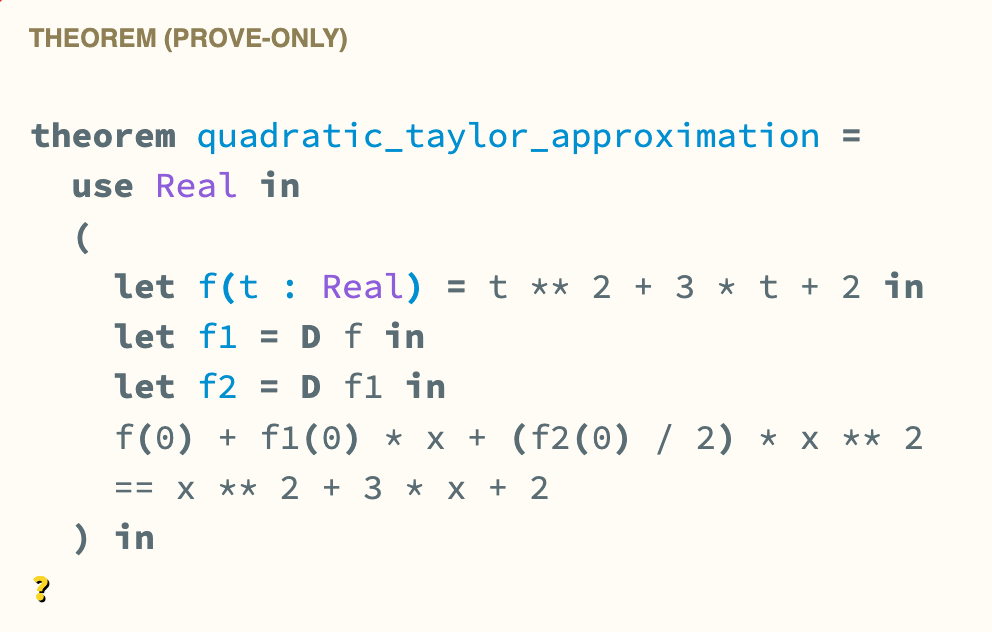}
    \caption{Dependent \texttt{let} bindings for quadratic Taylor approximation in Hazel Prover.
      The sequence defines $f$, computes $f_1$ from $f$, and then computes $f_2$ from $f_1$.}
    \label{fig:quadratic-taylor-let-chain}
    \Description{The theorem's dependent \texttt{let} bindings define a quadratic function f, define f1 as its derivative, and define f2 as the derivative of f1.}
  \end{figure*}

  \begin{figure*}[htbp]
    \centering
    \includegraphics[width=\textwidth]{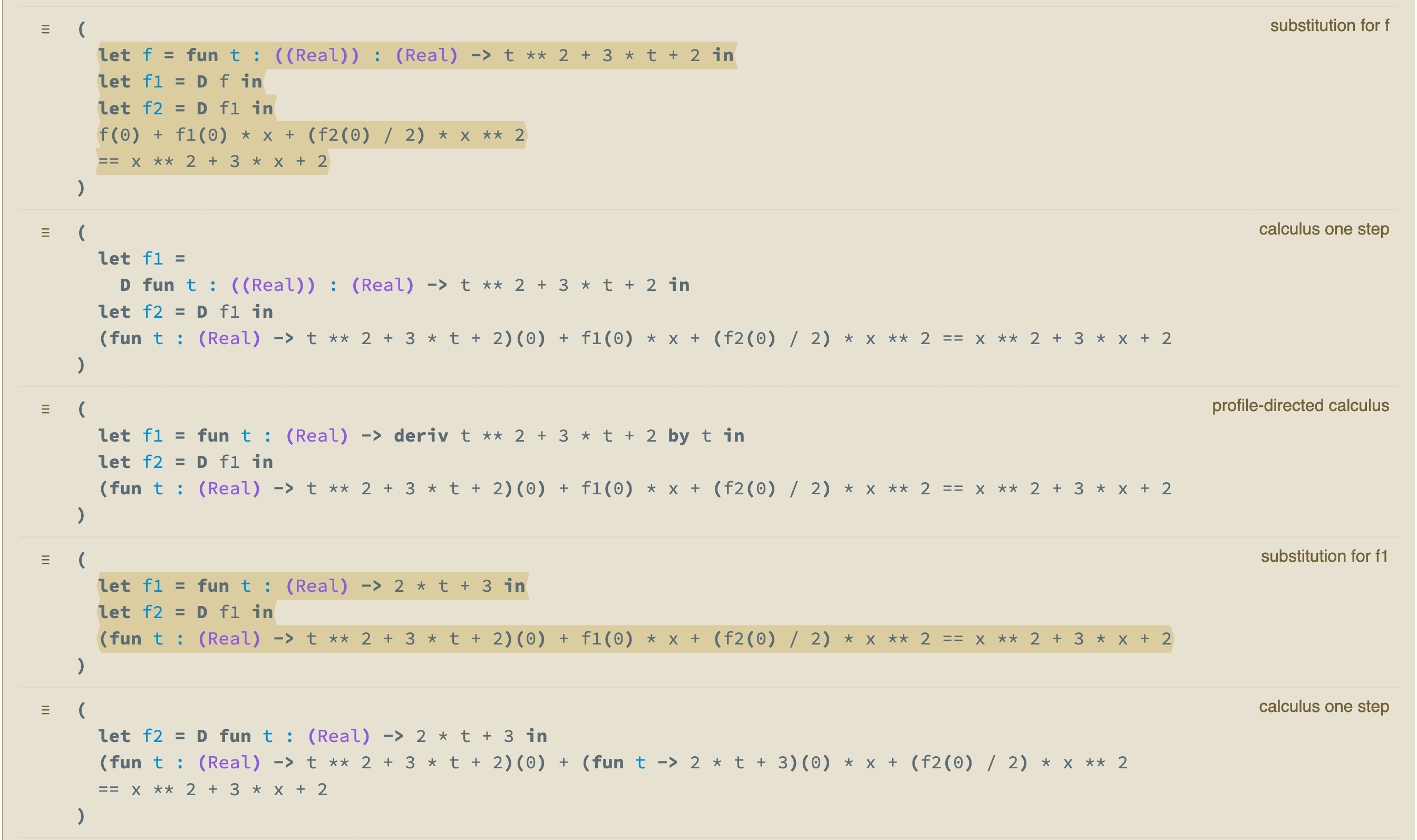}
    \caption{Derivative proof trace for quadratic Taylor approximation in Hazel Prover.
      The proof trace first differentiates $f(t)=t^2+3t+2$ to bind $f_1(t)=2t+3$, substitutes that computed function for $f_1$, and differentiates again to obtain $f_2(t)=2$.}
    \label{fig:quadratic-taylor-derivative-chain}
    \Description{A Hazel Prover proof trace: f is differentiated to compute f1, f1 is substituted into the next derivative, and that function is differentiated to compute f2.}
  \end{figure*}
}

\textbf{Trigonometric Taylor Approximation.}
The final case study asks users to compute a Taylor polynomial for the trigonometric function from the trigonometry case study, $\frac{7}{4} - \cos(2x) + \frac{1}{4}\cos(4x)$.
This more complex function is targeted at real-world numerical methods practitioners.
Though similar to the previous case study, the work is much more tedious due to the complexity of the function and number of chain rule derivatives, leading to high \textit{interaction overhead}.
This represents a failure mode of our interface at present; support for stepping and substituting subexpressions would reduce the tedium.
Rocq proof export does not yet support Taylor remainders, preventing formal proof export and limiting \textit{mathematical coverage}.


\newcommand{\showDOI}[1]{\unskip}
\bibliographystyle{ACM-Reference-Format}
\bibliography{sample-base}

\end{document}